\documentclass{ws-ijmpb}
\usepackage{mathrsfs,epsfig}

\newcommand\be{\begin{eqnarray}}
\newcommand\ee{\end{eqnarray}}

\begin{document}

\markboth{James M Lattimer and Madappa Prakash}
{What a Two Solar Mass Neutron Star Really Means}

%
\catchline{}{}{}{}{}
%

\title{What a Two Solar Mass Neutron Star Really Means}

\author{JAMES M. LATTIMER}

\address{Department of Physics and Astronomy\\
Stae University of New York at Stony Brook\\
Stony Brook, NY 11794-3800, USA\\
lattimer@mail.astro.sunysb.edu}

\author{MADAPPA PRAKASH}

\address{Department of Physics and Astronomy, Ohio University\\
Athens, Ohio 45701, USA\\
prakash@harsha.phy.ohiou.edu}

\maketitle

\begin{history}
\received{Day Month Year}
\revised{Day Month Year}
\end{history}

\begin{abstract}
The determination of neutron star masses is reviewed in light of a
new measurement of 1.97 M$_\odot$ for PSR J1614-2230 and an estimate
of 2.4 M$_\odot$ for the black widow pulsar.  Using a simple analytic
model related to the so-called maximally compact equation of state,
model-independent upper limits to thermodynamic properties in neutron
stars, such as energy density, pressure, baryon number density and
chemical potential, are established which depend upon the neutron star
maximum mass.  Using the largest well-measured neutron star mass, 1.97
M$_\odot$, it is possible to show that the energy density can never
exceed about 2 GeV, the pressure about 1.3 GeV, and the baryon
chemical potential about 2.1 GeV.  Further, if quark matter comprises
a significant component of neutron star cores, these limits are
reduced to 1.3 GeV, 0.9 GeV, and 1.5 GeV, respectively.  We also find
the maximum binding energy of any neutron star is about 25\% of the
rest mass.  Neutron matter properties and astrophysical constraints
additionally imply an upper limit to the neutron star maximum mass of
about 2.4 M$_\odot$.  A measured mass of 2.4 M$_\odot$ would be
incompatible with hybrid star models containing {\it significant}
proportions of exotica in the form of hyperons, Bose condensates or quark
matter.
\end{abstract}

\keywords{neutron star masses; equation of state.}

\section{Introduction}

One of the most fascinating stories in astrophysics concerns the
accumulation of precisely measured neutron star masses.  Gerry has had
a long-time interest in these measurements, as for many years he has
maintained that the neutron star maximum mass is no greater than
(1.5-1.6) M$_\odot$.  From a theoretical perspective, a maximum mass
of 1.5 M$_\odot$ fits his fondness for the effects of kaon
condensation proposed by Kaplan and Nelson~\cite{KN86}: the effective
kaon mass falls with increasing density and the eventual onset of kaon
condensation at a few times the nuclear saturation density softens the
equation of state (hereafter, EOS) and leads to a rather small maximum
mass.  He takes as observational support of this thesis the
``missing'' neutron star in the remnant of SN 1987A, which had a
baryon mass not in excess of 1.7 M$_\odot$ based on estimates of the
initial mass of the star and its ejected mass \cite{brown94}.  He
reasons that the maximum stable mass of the proto-neutron star as it
deleptonized must have eventually fallen below its actual mass,
prompting a collapse to a black hole\cite{thorsson94}.  Of course,
this had to have happened more than 12 seconds after the core collapse
since neutrinos were observed during that time period and their
emission is believed to abruptly cease~\cite{Burrows88} once the black
hole's event horizon forms.

\section*{Measurements of Neutron Star Masses}

The most accurate measurements of neutron star masses are for pulsars
in bound binary systems.  In these systems, five Keplerian parameters
can be precisely measured by pulse-timing techniques
\cite{manchester1977}, including the binary period $P$, the projection
of the pulsar's semimajor axis on the line of sight $a_p\sin i$ (where
$i$ is the binary inclination angle), the eccentricity $e$, and the
time and longitude of periastron $T_0$ and $\omega$.  Combining two of
the observational parameters, one can form the mass function:
\be 
f_p=\left({2\pi\over P}\right)^2\left({a_p\sin i\over
  c}\right)^3{{\rm~M}_\odot\over{\rm T}_\odot}={(M_c\sin i)^3\over
  M^2}{\rm~M}_\odot, 
\ee
where $M=M_p+M_c$ is the total mass, $M_p$ is the pulsar mass, and
$M_c$ is the companion mass (all measured in M$_\odot$ units).  The
constant T$_\odot=G{\rm M}_\odot/c^3$ is 4.9255 $\mu$ s.  The
mass function $f_p$ is also equal to the minimum possible mass $M_c$ for the
companion.

The inclination angle $i$ is often the most difficult aspect to infer,
but even if it was known {\it a priori} the above equation only
specifies a relation between $M_p$ and $M_c$ unless the mass function
$f_c$ of the companion is also measurable.  This occurs in the rare case
when the companion is itself a pulsar or a star with an observable
spectrum.

Fortunately, binary pulsars are compact systems and general
relativistic effects can often be observed.  These include the advance
of the periastron of the orbit 
\be
\dot\omega=3\left({2\pi\over P}\right)^{5/3}(M{\rm
  T}_\odot)^{2/3}(1-e^2)^{-1}, \ee 
the
combined effect of variations in the tranverse Doppler shift and
gravitational redshift around an elliptical orbit 
\be
\gamma=e\left(\frac{P}{2\pi}\right)^{1/3}\frac{M_c(M+M_c)}{M^{4/3}}{\rm
  T}_\odot^{2/3}, 
\ee 
and
the orbital period decay due to the emission of gravitational
radiation 
\be 
\dot P=-{192\pi\over5}\left({2\pi{\rm T}_\odot\over
P}\right)^{5/3}\left(1+{73\over24}e^2+{37\over96}e^4\right)
(1-e^2)^{-7/2}{M_pM_c\over M^{1/3}}.  
\ee 
The inclination angle can be constrained by measurements of two or
more of these effects.  However, only in extremely compact systems is
this precisely possible.  Otherwise, additional effects, such as an
eclipse or limits obtained from the lack of an eclipse, or Shapiro
time delay, which is caused by the propagation of the pulsar signal
through the gravitational field of its companion, must be observed.
The Shapiro~\cite{Shapiro64} delay in general relativity produces a
delay in pulse arrival time~\cite{damour86,freirewex10}
\be 
\delta_S(\phi)=2M_c{\rm
T}_\odot\ln\left[{1+e\cos\phi\over1-\sin(\omega+\phi)\sin i}\right],
\ee 
where $\phi$ is the true anomaly, the angular parameter defining the
position of the pulsar in its orbit relative to the periastron.  $\delta_S$ 
is a periodic function of $\phi$ with an approximate amplitude,
\be
\Delta_S\simeq2M_cT_\odot\left|\ln\left[\left({1+e\sin\omega\over1-e\sin\omega}\right)\left({1+\sin i\over1-\sin i}\right)\right]\right|,
\label{ds}
\ee 
which is large only for edge-on binaries, $\sin i\sim1$ or those which have
both large eccentricities and large magnitudes of $\sin\omega$.  Only a
fraction of pulsars in binaries have two or more sufficiently
well-measured relativistic effects to enable precise measurements of
the pulsar mass $M_p$.

In some cases, the companion star can be optically detected.  In most
cases, the companion is a white dwarf.  It is possible to estimate
white dwarf masses from observations of the optical flux, effective
temperature and distance, with the latter being especially
problematic.  Another possibility is to measure the surface gravity
from spectral measurements, but currently this is subject to 
large systematic effects.

Less accurate measurements of neutron star masses are possible in
X-ray binaries, in which X-rays are emitted by matter accreting onto a
neutron star from a companion star. Both X-ray and optical
observations can yield both mass functions $f_p$ and $f_c$, but $f_p$
is subject to large uncertainties due to the faintness of optical
radiation.  In the case of the black widow pulsar, PSR B1957+20,
optical observations of the companion yield both a mass function and
an estimate of the inclination $i$ from the shape of the light curve
\cite{reynolds2007}.
The inferred semimajor axis of the companion's orbit has to be
corrected for the finite size of the companion, however (see below),
introducing a source of systematic error in the pulsar mass
determination of this system.  Nevertheless, interesting lower limits to
the pulsar mass are obtained \cite{vankerkwijk10}.  

\section*{Recent Measurements}

\begin{figure}[bt]
\centerline{\psfig{file=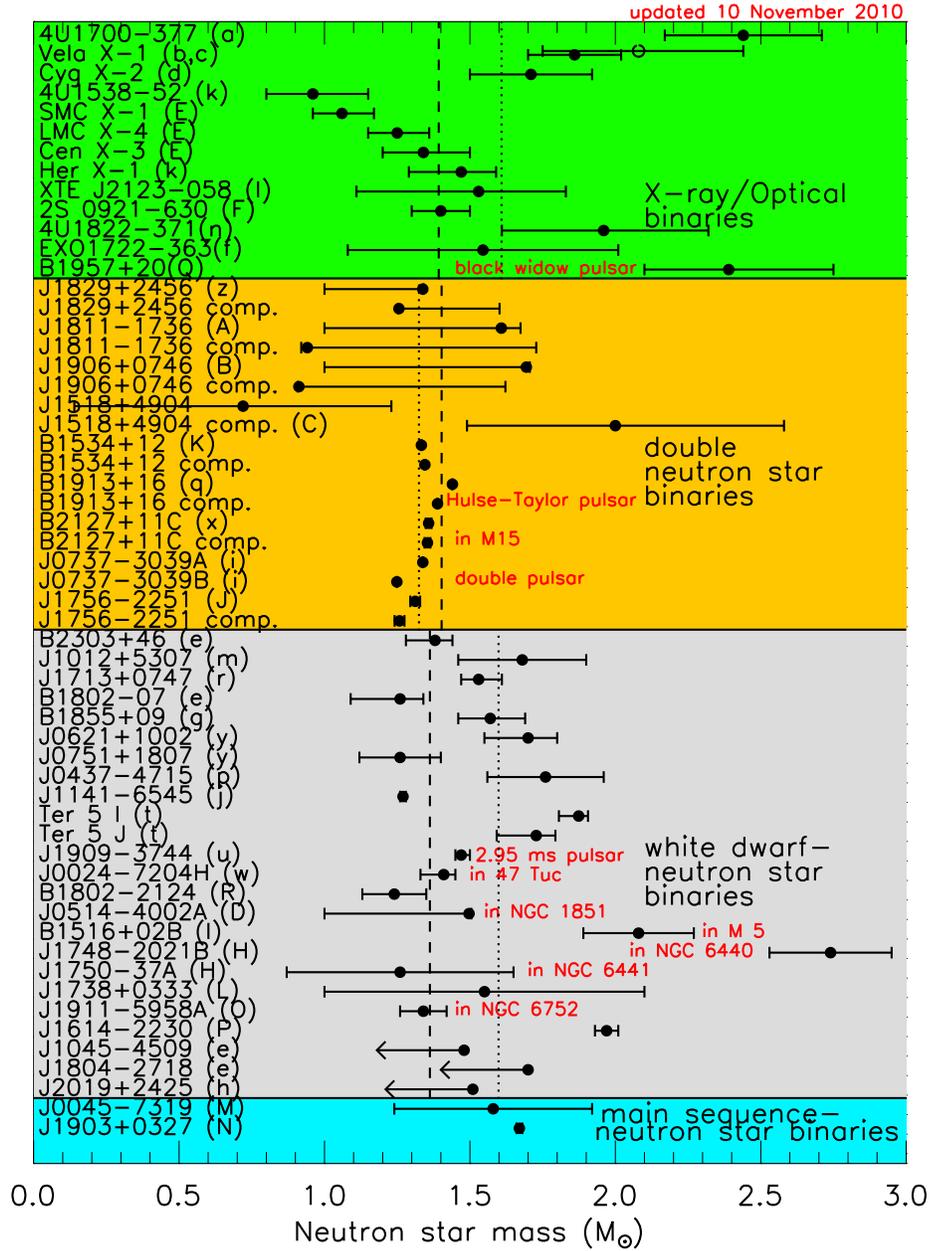,width=5.75in}}
\caption{Measured neutron star masses. References in parenthesis following source numbers are identified in Table \ref{masses}.}
\label{masses}
\end{figure}

The most recent information, as of November 2010,
is summarized in Figure \ref{masses} and Table \ref{mass} and
collects information from Refs. \refcite{clark02} to
\refcite{ferdman10}.  This compilation represents a significant update
to the figure of observed masses and references in Ref. \refcite{LP05}.

\begin{table}
\tbl{Neutron star mass measurements with
$1\sigma$ uncertainties.  Reference letters correspond to Fig. 1.
For each group, mean and weighted masses are indicated.} 
{\begin{tabular}{ ccccccc}
\hline\Hline \\[-1.7ex]
Object & Mass (M$_\odot$) & Reference & \qquad &
Object & Mass (M$_\odot$) & Reference \\[1ex]
\hline \\[-1.7ex]
\multicolumn{7}{c}{\it X-Ray/Optical Binaries (mean $=1.609$ M$_\odot$, weighted mean $=1.393$ M$_\odot$)}\\[1ex]

4U1700-377 & $2.44^{+0.27}_{-0.27}$ & a (\refcite{clark02}) & \qquad  & 
Vela X-1 & $1.86^{+0.16}_{-0.16}$ & b, c (\refcite{barziv01,quaintrell03}) \\

Cyg X-2 & $1.71^{+0.21}_{-0.21}$ & d (\refcite{casares10}) & & 
4U1538-52 & $0.96^{+0.19}_{-0.16}$ & k (\refcite{vankerkwijk95}) \\

SMC X-1 & $1.06^{+0.11}_{-0.10}$ & E (\refcite{vandermeer07}) & & 
LMC X-4 & $1.25^{+0.11}_{-0.10}$ & E (\refcite{vandermeer07}) \\

Cen X-3 & $1.34^{+0.16}_{-0.14}$ & E (\refcite{vandermeer07}) & & 
Her X-1 & $1.47^{+0.12}_{-0.18}$ & k (\refcite{vankerkwijk95}) \\

XTE J2123-058 & $1.53^{+0.30}_{-0.42}$ & l (\refcite{gelino03,tomsick04}) & & 
2S 0921-630 & $1.4^{+0.1}_{-0.1}$ & F (\refcite{steeghs07}) \\

4U 1822-371 & $1.96^{+0.36}_{-0.35}$ & n (\refcite{jonker03}) & &
EXO 1722-363 & $1.545^{+0.465}_{-0.465}$ & f (\refcite{mason10})\\

B1957+20 & $2.39^{+0.36}_{-0.29}$ & Q (\refcite{vankerkwijk10}) & & &
& \\[1ex]
\hline \\[-1.7ex]

\multicolumn{7}{c}{\it Neutron Star -- Neutron Star Binaries (mean $=1.325$ M$_\odot$, weighted mean $=1.403$ M$_\odot$)}\\[1ex]

J1829+2456 & $1.338^{+0.002}_{-0.338}$ & z (\refcite{champion05}) & &
J1829+2456 (c) & $1.256^{+0.346}_{-0.003}$ & z (\refcite{champion05}) \\

J1811-1736 & $1.608^{+0.066}_{-0.608}$ & A (\refcite{corongiu04}) & &
J1811-1736 (c) & $0.941^{+0.787}_{-0.021}$ & A (\refcite{corongiu04}) \\

J1906+07 & $1.694^{+0.012}_{-0.694}$ & B (\refcite{lorimer06}) & &
J1906+07 (c) & $0.912^{+0.710}_{-0.004}$ & B (\refcite{lorimer06}) \\

J1518+4904 & $0.72^{+0.51}_{-0.58}$ & C (\refcite{janssen08}) & &
J1518+4904 (c) & $2.00^{+0.58}_{-0.51}$ & C (\refcite{janssen08}) \\

1534+12 & $1.3332^{+0.0010}_{-0.0010}$ & K (\refcite{stairs02}) & & 
1534+12 (c) & $1.3452^{+0.0010}_{-0.0010}$ & K (\refcite{stairs02}) \\

1913+16 & $1.4398^{+0.0002}_{-0.0002}$ & q (\refcite{weisberg10}) & &
1913+16 (c) & $1.3886^{+0.0002}_{-0.0002}$ & q (\refcite{weisberg10}) \\

2127+11C & $1.358^{+0.010}_{-0.010}$ & x (\refcite{jacoby06}) & &
2127+11C (c) & $1.354^{+0.010}_{-0.010}$ & x (\refcite{jacoby06}) \\

J0737-3039A & $1.3381^{+0.0007}_{-0.0007}$ & i (\refcite{kramer06}) & & 
J0737-3039B & $1.2489^{+0.0007}_{-0.0007}$ & i (\refcite{kramer06}) \\

J1756-2251 & $1.312^{+0.017}_{-0.017}$ & J (\refcite{ferdman08}) & &
J1756-2251 (c) & $ 1.258^{+0.017}_{-0.017}$ & J (\refcite{ferdman08}) \\[1ex]
\hline \\[-1.7ex]
\multicolumn{7}{c}{\it Neutron Star -- White Dwarf Binaries (mean $=1.599$ M$_\odot$, weighted mean $=1.362$ M$_\odot$)}\\[1ex]

B2303+46 & $1.38^{+0.06}_{-0.10}$ & e (\refcite{thorsett99}) & & 
J1012+5307 & $1.68^{+0.22}_{-0.22}$ & m (\refcite{lange01}) \\

J1713+0747 & $1.53^{+0.08}_{-0.04}$ & r (\refcite{splaver05,tauris99}) & & 
B1802-07 & $1.26^{+0.08}_{-0.17}$ & e (\refcite{thorsett99}) \\

B1855+09 & $1.57^{+0.12}_{-0.11}$ & g (\refcite{nice03b,tauris99}) & & 
J0621+1002 & $1.70^{+0.10}_{-0.17}$ & y (\refcite{nice08}) \\

J0751+1807 & $1.26^{+0.14}_{-0.14}$ & y (\refcite{nice08}) & & 
J0437-4715 & $1.76^{+0.20}_{-0.20}$ & p (\refcite{verbiest08}) \\

J1141-6545 & $1.27^{+0.01}_{-0.01}$ & j (\refcite{bhat08}) & & 
Ter 5 I & $1.874^{+0.32}_{-0.068}$ & t (\refcite{ransom05}) \\

Ter 5 J & $1.728^{+0.066}_{-0.136}$ & t (\refcite{ransom05}) & &
J1909-3744 & $1.47^{+0.03}_{-0.02}$ & u (\refcite{hotan06}) \\

J0024-7204H & $1.41^{+0.04}_{-0.08}$ & w (\refcite{freire03}) & &
B1802-2124 & $1.24^{+0.11}_{-0.11}$ & R (\refcite{ferdman10}) \\

J0514-4002A & $ 1.497^{+0.008}_{-0.497}$ & D (\refcite{freire07}) & &
B1516+02B & $2.08^{+0.19}_{-0.19}$ & I (\refcite{freire08a}) \\

J1748-2021B & $2.74^{+0.21}_{-0.21}$ & H (\refcite{freire08b}) & &
J1750-37A & $1.26^{+0.39}_{-0.39}$ & H (\refcite{freire08b}) \\

J1738+0333 & $1.55^{+0.55}_{-0.55}$ & L (\refcite{freire08c}) & &
J1911-5958A & $1.34^{+0.08}_{-0.08}$ & O (\refcite{bassa06}) \\

J1614-2230 & $1.97^{+0.04}_{-0.04}$ & P (\refcite{demorest10}) & &
J1045-4509 & $<1.48$ & e (\refcite{thorsett99}) \\

J1804-2718 & $<1.70$ & e (\refcite{thorsett99}) & &
J2019+2425 & $<1.51$ & h (\refcite{nice01}) \\[1ex]
\hline \\[-1.7ex]
\multicolumn{7}{c}{\it Neutron Star -- Main Sequence Binaries}\\[-.3ex]
J0045-7319 & $1.58^{+0.34}_{-0.34}$ & M (\refcite{nice03a}) & &
J1903+0327 & $1.67^{+0.01}_{-0.01}$ & N (\refcite{freire10}) \\[1ex]
\hline\Hline
\end{tabular}}
\label{mass}
\end{table}

There is now ample observational support from pulsars for neutron
stars with masses significantly greater than 1.5 M$_\odot$.  These
include PSR J1903+0327 which has a main-sequence companion; the
pulsars I and J in the globular cluster Ter 5, the globular cluster
binaries PSR J1748-2021B and PSR B1516+02B, and PSR J1614-2230, all
with white-dwarf companions; and the black widow pulsar B1957+20.  The
inclination angles of the two binaries containing pulsars I and J in the
globular cluster Ter 5 are unconstrained by observation, but if their
inclinations are individually randomly directed, there is a 95\%
chance that at least one of these pulsars is greater than 1.68
M$_\odot$ \cite{ransom05}.  Assumptions related to random orientation
also have roles in the mass determinations for PSR J1748-2021B and PSR
B1516+02B.  One has to be careful about the assumption of random
orientation, however, as the selection of potential candidates for
follow-up observations necessary for accurate mass determinations are
biased.  The case of PSR J0751+1807 illustrates the point: initially
estimated to have a mass $2.2\pm0.2$ M$_\odot$, the mass was recently
established to be 1.26 M$_\odot$ upon measurement of another
relativistic parameter.  Nevertheless, about 2 years ago, a
measurement of the inclination of PSR J1903+0327 \cite{freire10} using
Shapiro time delay lead to an accurately determined mass of
$1.67\pm0.01$ M$_\odot$, replacing PSR B1913+16 as establishing the
value for the minimum maximum mass.  

But even more outstanding is the recent determination of $1.97\pm0.04$
M$_\odot$ for the mass of PSR J1614-2230\cite{demorest10}, also using
Shapiro time delay to measure the inclination.  This 3.15 ms pulsar is
in an 8.7 d nearly circular orbit about an 0.5 M$_\odot$ companion,
with $a_p\sin i=11.3$ light-second and $\sin i=0.99989$, {\it i.e.},
it is almost edge on.  The Shapiro time delay amplitude, from
Eq. ({\ref{ds}), becomes 
\be
\Delta_s\simeq4M_cT_\odot\ln(2/\cos i)\simeq 48.8~ \mu{\rm s}\,.
\ee
By virtue of its accuracy, this mass
measurement has become the new standard for the minimum value of the
neutron star maximum mass.

In addition, a number of X-ray binaries seem to contain
high-mass neutron stars: about 1.9 M$_\odot$ in the case of Vela X-1
and 2.4 M$_\odot$ in the case of 4U 1700-377.  Nonetheless, the large
systematic errors inherent in X-ray binary mass measurements warrant
caution.

The black widow pulsar represents an intriguing case, with a best
estimate of about 2.4 M$_\odot$.  This system has both pulsar timing
and optical light curve information.  It consists of a 1.6 ms pulsar
in a nearly circular 9.17~h orbit around an extremely low mass
companion, $M_c\simeq0.03$ M$_\odot$.  The pulsar is eclipsed for
about 50--60 minutes of each orbit, but considering that $a_p\sin
i=0.089$ light-second = 0.038 R$_\odot$, and $a_c\sin i\sim3$
R$_\odot$ is $M_p/M_c\simeq80$ times larger, the eclipsing object has
to be much larger ($\sim3$ R$_\odot$/10) than the size of the
companion star.  It is believed that irradiation of the companion by
the pulsar strongly heats its nearside to the point of ablation,
leading to a comet-like tail and a large cloud of plasma which is
responsible for the eclipsing.  The pulsar is literally consuming its
companion, hence the name ``black widow'', and has reduced its mass to
a small fraction of its original mass.  The irradiation also produces
an enormous (factor of 100) variation in the brightness of the
companion depending on how much of its nearside is visible during its
orbit.  The companion is bloated and nearly fills its Roche lobe.  The
companion's optical light curve allows the mass ratio $M_p/M_c$ and
the inclination angle $i$ to be estimated.  However, the large size of
the companion means that the ``center of light'' of the system is not
equivalent to its ``center of mass'': the optical light curve depends
on the projected semi-major axis of the irradiated nearside of the
companion rather than the projected semi-major axis of the center of
mass of the companion.  The extreme cases are either that the
companion has zero radius or that it completely fills its Roche lobe,
but estimates based on modeling considerably reduce the allowed range.
The extremes lead to a range $1.7<M_p/{\rm M}_\odot<3.2$, but the
likely value is $~2.4\pm0.4$ M$_\odot$.  It will clearly be valuable
to extend observations and modeling of this system since a 2.4
M$_\odot$ neutron star would have profound implications.

\section*{What Gerry Would Say}

We would love to be able to argue with Gerry about these observations
and their implications, 
and so we felt this was an ideal topic to explore in
this contribution.  We can, however, guess what Gerry would have said:
``We can't tell nature how to behave; we have to explain it'' and, as
he often quoted Bismarck, ``Was schert mich mein Geschwaetz von
gestern?''  (``Why do I care about my twaddle from yesterday?'')

\section*{What We Say}

One of the most surprising consequences of a large pulsar mass
measurement is that it severely restricts values of the maximum
central density and pressure in that star, and, by extension, that of
any other neutron star as well.  This realization was elaborated in
Ref. \refcite{LP05}, but it can be extended to restrict the
possibilities for the presence of exotic matter, such as hyperons,
Bose condensates and/or quark matter, in neutron star interiors.
Moreover, we show here that upper limits to the chemical potentials of
quark or baryonic matter can be established using the available mass
data, and that these are smaller than often employed in calculations
of the properties of quark matter.  The consequences of
well-determined large neutron star masses is the topic to which we now
turn our attention.

\section{Neutron Star Structure}

The global properties of neutron stars are determined, assuming general
relativity, by the 
Tolman-Oppenheimer-Volkov (TOV) equations~\cite{TOV}
\be
{dp\over dr} &=& -{G(\varepsilon+p)(m+4\pi r^3p/c^2)\over
r(rc^2-2Gm)}\,, \label{tov0} \\
{dm\over dr} &=& 4\pi r^2\varepsilon/c^2\,, \label{tov}\\
{d{\cal N}\over dr} &=& 4\pi r^2 n \left(1-{2Gm\over rc^2}\right)^{-1/2}\,, \label{tov1}
\ee
\begin{figure}[bt]
\begin{minipage}[l]{0.495\linewidth}
\psfig{file=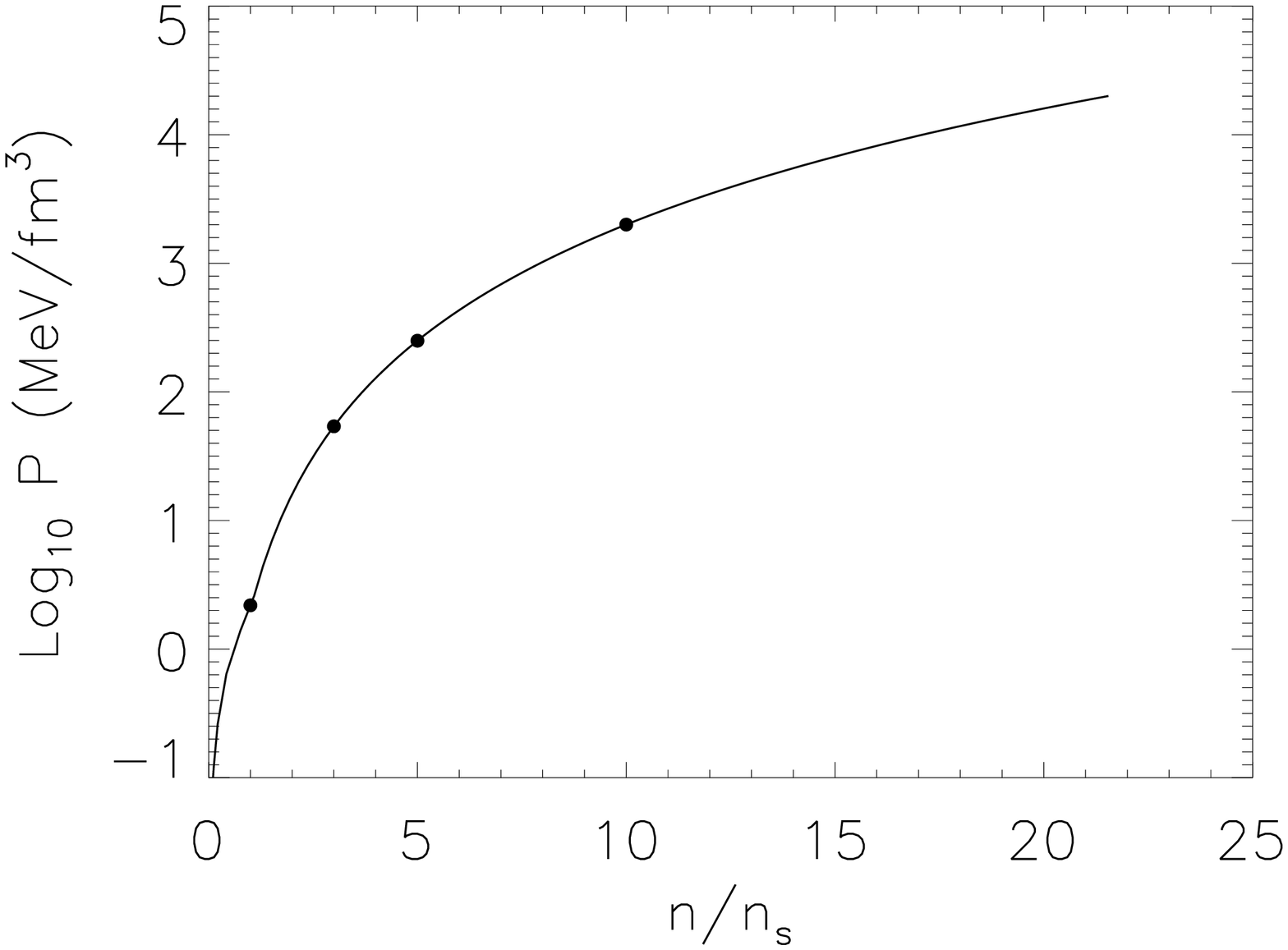,angle=90,width=2.9in}
\end{minipage}
\begin{minipage}[r]{0.495\linewidth}
\psfig{file=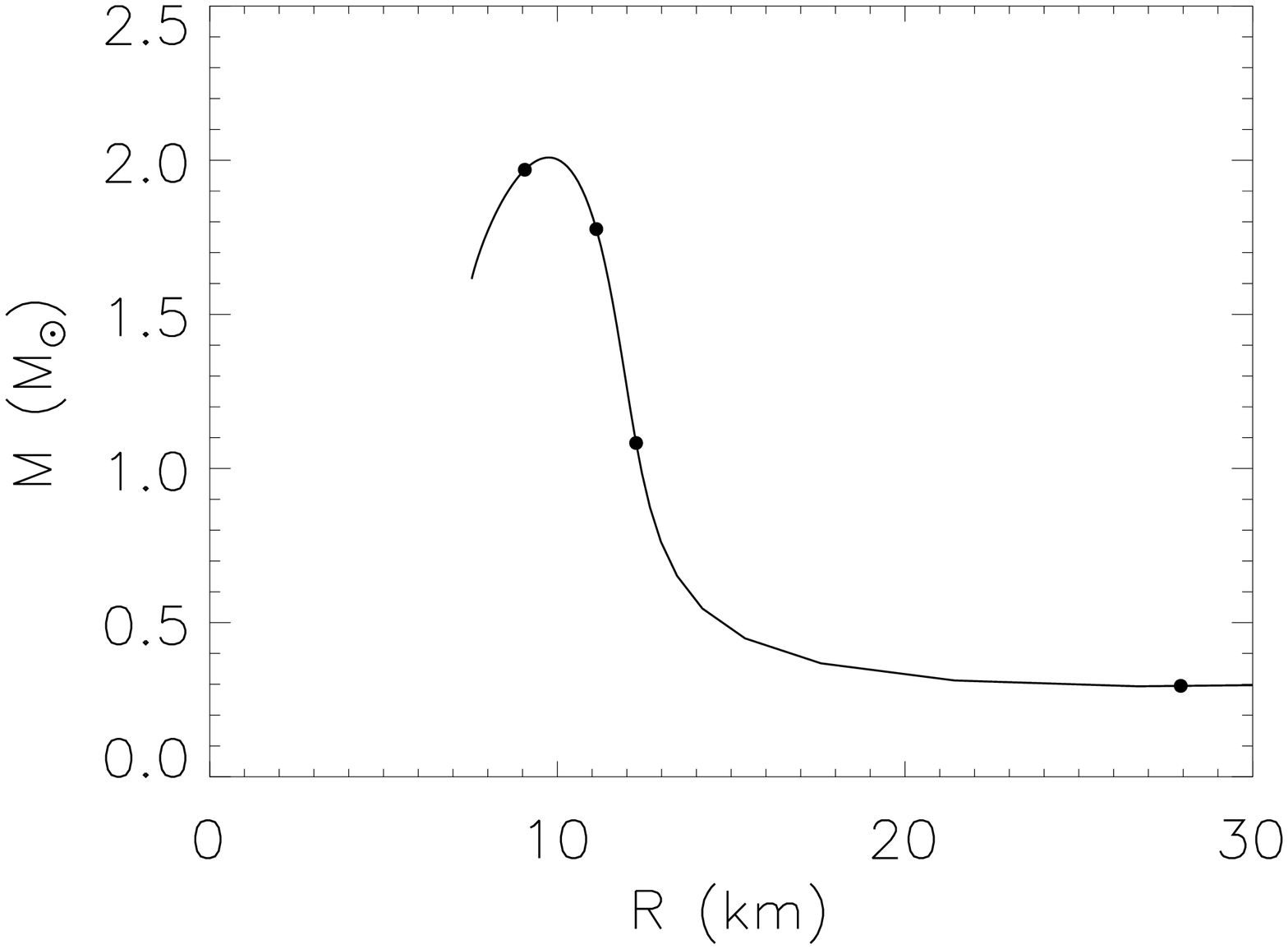,angle=90,width=2.9in}
\end{minipage}
\caption{Solution of the TOV equations.  Integrations beginning at the
star's center with pressures and baryon number densities (in units of
the nuclear saturation density $n_s$) shown by the dots in the left
panel result in specific masses and radii, shown by the corresponding
dots in the right panel.}
\label{tovfig}
\end{figure}
\noindent where $m$ is the enclosed gravitational mass, ${\cal N}$ is
the enclosed baryon number, $p$ is the pressure, $\varepsilon$ is the
total energy density, $n$ is the baryon number density, and $r$ is the
radial variable.  Choosing a value of the pressure or baryon or energy
density at the star's center, and integrating to the surface where the
pressure vanishes, results in a specific mass and radius, as shown in
Fig. \ref{tovfig}.  A consequence of general relativity is the
existence of a maximum mass: configurations with larger central
pressures are dynamically unstable and collapse into black holes.

It is convenient to recast the TOV equations in terms of the
dimensionless variable $h$ defined by $dh=dp/(\varepsilon+p)=d\mu/\mu$
where $\mu=(\varepsilon+p)/n$ is the baryon chemical potential, which
in beta equilibrium is the same as the neutron chemical potential.
Therefore,
\be
\mu=\mu_oe^h\,,
\label{muh}
\ee
where $\mu_o=\varepsilon_o/n_o$, $\varepsilon_o$ and $n_o$ are the
chemical potential, energy density and baryon number density,
respectively, where $p=h=0$.  For a star with a normal crust,
$\mu_o\simeq930$ MeV, the approximate mass-energy of iron nuclei per
baryon. Note that the quantity $h$ is related to the
metric function $\nu$, where $g_{tt}=e^\nu$ and $dh=-d\nu/2$.  One finds
\be
{dr^2\over dh}&=&-\frac{2r^2c^2}{G} \left[{rc^2-2Gm\over mc^2+4\pi r^3p}\right]
\,, \label{TOV1}\\
{dm\over dh}&=&-{4\pi r^3\varepsilon\over G} \left[{rc^2-2Gm\over mc^2+4\pi r^3p}\right]\,,\label{TOV2}\\
{d{\cal N}\over dh}&=&-{4\pi r^{7/2}nc^3\over G} \left[{\sqrt{rc^2-2Gm}\over mc^2+4\pi r^3p}\right]\,.\label{TOV3}
\ee
Writing the TOV equations this way ensures that near both the
center and the surface the right-hand sides are finite.  The equations
are integrated from a central value $h=h_c$ to $h=0$.  In order to
obtain a specific total gravitational mass
$M=m(h=0)$, one iterates to find the appropriate value of $h_c$.  For a
given EOS, that is, the functions $\varepsilon(h)$ and $p(h)$, a
specific value or $h_c$, $h_{max}$, results in the maximum mass configuration
where $M=M_{max}, N=N_{max}$ and $R=R_{max}$ where $R=r(h=0)$ is the
stellar radius and $N={\cal N}(h=0)$ is the total baryon number.
The baryon number is useful in the definition of the binding energy
\be
BE = (Nm_b-M)c^2\,,\label{be}
\ee
where $m_b$ is the baryon mass.

It is an inescapable consequence of general relativity that all EOSs
result in a maximum mass configuration, in contrast to Newtonian
gravity for which no such limit exists.  It is intuitive that the
largest maximum masses will be a consequence of the {\it stiffest}
EOSs, {\it i.e.}, those with the largest pressures for a given
density.  It is commonly assumed that the maximum realistic limit to
the pressure is obtained by assuming causality, in other words, that
the sound speed is limited everywhere by light speed.  The assumption
of causality led Rhoades and Ruffini \cite{RR74} to their estimate for
the maximum neutron star mass.  It is logical, therefore, that the
maximum mass configuration so obtained must also have the largest
possible values for the central pressure, energy density and chemical
potential.  The main motivation of this paper is to explore the
consequences of this in light of new observations of large neutron
star masses.

It was argued by Haensel and Zdunik~\cite{haensel89} and by Koranda,
Stergioulas and Friedman~\cite{KSF97} that the most compact
configurations result from the combination of a soft EOS at small
densities and a stiff EOS at large densities.  The extreme limit of
this situation is shown in Fig. \ref{compfig}, where the pressure
vanishes below the density $\varepsilon_o$ and is assumed causal at
{\it all} higher densities (the so-called maximally compact EOS):
\begin{figure}[bt]
\centerline{\psfig{file=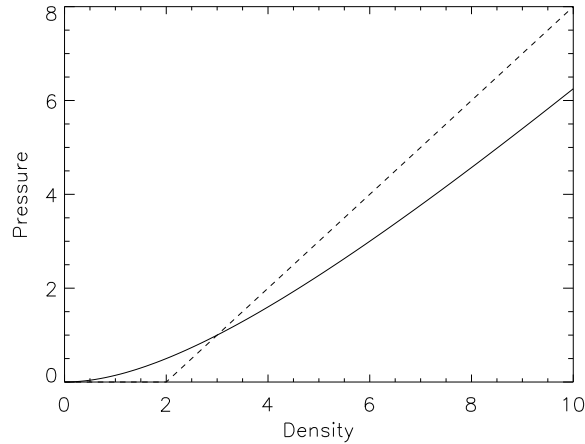,angle=90,width=3.5in}}
\caption{The solid line is a typical neutron star EOS.  The dashed
line shows the maximally compact EOS formed by minimizing the pressure
$p$ at low energy density $\varepsilon$ ({\it i.e.}, $p=0$) and
maximizing $p$, subject to causality, at high density ({\it i.e.},
$p=\varepsilon-\varepsilon_o$).  $p$ and $\varepsilon$ are in the same
arbitrary units, with $\varepsilon_o$ in this example set to 2.}
\label{compfig}
\end{figure}
\be
p=0\qquad \varepsilon\le\varepsilon_o, \hspace*{2cm} p=\varepsilon-\varepsilon_o \qquad \varepsilon\ge\varepsilon_o. \label{mc} 
\ee
Since $\varepsilon_o$ is the only EOS parameter,  
the TOV equations scale with it according to~\cite{Witten84}
\be
\varepsilon\propto\varepsilon_o,\qquad p\propto\varepsilon_o,\qquad
n\propto\varepsilon_o,\qquad 
m\propto\varepsilon_o^{-1/2},\qquad r\propto\varepsilon_o^{-1/2}\,.
\label{scale}
\ee
Following Lattimer, Prakash, Masak and Yahil~\cite{LPMY90} (hereafter LPMY), 
we wish to study a somewhat more general EOS, which encompasses
 the maximally compact EOS as a special case,  namely,
\be
p=s(\varepsilon-\varepsilon_o)\,,
\label{spec}
\ee where $s$ is a constant representing the square of the adiabatic
sound speed. We can then express $\varepsilon$, $p$ and the baryon number
density $n$ in terms of $h$
as
\be
\varepsilon=t^{-1}\varepsilon_o(s^{-1}e^{ht}+1) \,, \qquad
p=t^{-1}\varepsilon_o(e^{ht}-1) \,,\qquad n=\varepsilon_oe^{h/s}/\mu_o \,
\label{caus}
\ee
where $t=1+1/s$. Using the dimensionless variables
\be
x=r^2\epsilon_oG/c^4\,,\qquad y=m\epsilon_o^{1/2}G^{3/2}/c^4\,,\qquad
z={\cal N}\epsilon_o^{1/2}\mu_oG^{3/2}/c^6
\label{xy}
\ee
the scale-free TOV equations become
\begin{eqnarray}
{dx\over dh}&=&-{2x(x^{1/2}-2y)\over y+4\pi x^{3/2}p/\varepsilon_o}\,,\label{TOV4}\\ 
{dy\over dh}&=&2\pi x^{1/2}{\varepsilon\over\varepsilon_o}{dx\over dh}\,,\label{TOV5}\\
{dz\over dh}&=&2\pi{n\mu_o\over\varepsilon_o}{x\over(x^{1/2}-2y)^{1/2}}{dx\over dh}\,.\label{TOV6}
\end{eqnarray}

For a given value of $s$, the value of $h_{max}$ is the value of $h_c$
that maximizes the total gravitational mass $y(h=0)$.  By virtue of
Eq. (\ref{muh}), $h_{max}$ then determines the maximum possible
chemical potential for that EOS.  Table \ref{table1} summarizes the
properties of the maximum mass solutions for the cases $s=1, 4/5, 2/3$
and $s=1/3$.  The case $s=1$ represents the maximally compact star; as
discussed in \S 4, the case $s=1/3$ is potentially applicable to quark
stars or stars with quark cores. For each $s$ are listed the central
values of $h$, $\varepsilon$ and $p$, and the surface values of $x, y$
and $z$, together with other dimensionless quantities of interest.

We will demonstrate that the maximally compact and the $s=1/3$ EOSs
result in useful approximate limits for the maximum values of
thermodynamic quantities in the interior of neutron and quark stars as
well as their maximum masses.

\section{Maximally Compact EOS}

We first examine the maximally compact EOS, {\it i.e.}, the case
$s=1$.  Expressing the results for $s=1$ shown in Table \ref{table1}
in astrophysical units (note that 1 MeV fm$^{-3}=1.323\times10^{-6}$
km$^{-2}$, and 1 M$_\odot=1.477$ km), we find
\be
M_{max} &=& y_{max}c^4\epsilon_o^{-1/2}G^{-3/2}=4.09~(\varepsilon_s/\varepsilon_o)^{1/2}{\rm~M}_\odot \,, \nonumber \\
N_{max}m_b &=& z_{max}m_bc^6\epsilon_o^{-1/2}\mu_o^{-1}G^{-3/2}=5.41~({m_bc^2/\mu_o})(\varepsilon_s/\varepsilon_o)^{1/2}{\rm~M}_\odot \,, \nonumber \\
R_{max} &=& x_{max}c^2\epsilon_o^{-1/2}G^{-1/2}=17.07~(\varepsilon_s/\varepsilon_o)^{1/2}{\rm~km} \,,\nonumber \\
\varepsilon_{max} &=& (e^{2h_{max}}+1)\varepsilon_o/2=3.034~\varepsilon_o\,,\nonumber \\
p_{max} &=& (e^{2h_{max}}XS-1)\varepsilon_o=2.034~\varepsilon_o\,,\nonumber \\
\mu_{max} &=& \mu_oe^{h_{max}}=2.251~\mu_o\,, \nonumber \\
n_{max} &=& (\varepsilon_{max}+p_{max})e^{-h_{max}}/\mu_o=
2.251~(\varepsilon_o/\mu_o) \,.
\label{mr} 
\ee
where $\varepsilon_s\simeq m_bn_s\simeq150$ MeV is the mass-energy
density at normal nuclear saturation baryon density $n_s\simeq0.16$
fm$^{-3}$.  The result for $M_{max}$ is very close to, but slightly
smaller than, that discussed by Rhoades \& Ruffini~\cite{RR74}, the
difference being their substitution of a low-density neutron star
crust EOS for densities less than $\varepsilon_f$, where
$\varepsilon_f>\varepsilon_o$ and $p_f>0$.  However, the difference is
very small, showing the validity of the maximally compact EOS for an
approximate limit.
\begin{table}[ph]
\caption{Dimensionless TOV eigenvalues for the maximum mass configurations 
for different $s$ values.  The case marked $1/3^*$ has $a_4=1/2$ and $a_2^{1/2}/B_{eff}^{1/4}=2.143$.  The notation $^{**}$ denotes self-bound
stars with hadronic shells as described in the text.
$h_{max},\varepsilon_{max}$ and $p_{max}$ are central values;
$x_{max}, y_{max}$ and $z_{max}$ are surface values.}
{\begin{tabular}{|c|c|c|c|c|c|c|c|c| }
\hline\Hline \\[-2.8ex]
$s$ & $h_{max}$ & $\varepsilon_{max}/\varepsilon_o$ & $p_{max}/\varepsilon_o$ & $x_{max}$ &$y_{max}$ & $z_{max}$ & $\varepsilon_{max}y_{max}^2/\varepsilon_o$& $x_{max}^{1/2}/y_{max}$ \\ [0.8ex] 
\hline \\[-2.8ex]
1& 0.8116 & 3.034 & 2.034 & 0.05779 & 0.08513 & 0.1127 & 0.02199 & 2.824 \\ 
$1^{**}$& 0.855 & 3.00 & 2.00 & 0.0635 & 0.0854 & 0.109 & 0.0219 & 2.95 \\
4/5& 0.727 & 3.29 & 1.83 & 0.0535 & 0.0787 & 0.103 & 0.0204 & 2.94 \\  
2/3& 0.663 & 3.55 & 1.70 & 0.0500 & 0.0732 & 0.0946 & 0.0190 & 3.05 \\  
1/2& 0.569 & 4.01 & 1.51 & 0.0443 & 0.0644 & 0.0810 & 0.0166 & 3.27 \\  
1/3 &  0.4520 & 4.826 & 1.275 & 0.03649 & 0.05169 & 0.06212 & 0.01289 & 3.696 \\ 
$1/3^*$ & 0.4350 & 6.585 & 1.639 & 0.02696 & 0.04307 & 0.05134 & 0.01221 & 3.813 \\
$1/3^{**}$ &  0.491 & 4.68 & 1.23 & 0.0439 & 0.0522 & 0.0607 & 0.0128 & 4.01 \\  
\hline\Hline
\end{tabular}}
\label{table1}
\end{table}

The results for $\varepsilon_{max}/\varepsilon_o$ and
$p_{max}/\varepsilon_o$ for the case $s=1$ were previously obtained by
Ref.~\refcite{KSF97}.  What they did not apparently realize was the
significance of the quantity
$\varepsilon_{max}y_{max}^2/\varepsilon_o$, which, when expressed in
astrophysical units for the maximally compact case $s=1$, is
\be 
\varepsilon_{max}M_{max}^2=1.358\times10^{16}{\rm~g~cm}^{-3}{\rm~M}_\odot^2\,.
\label{dm2}
\ee 
This equation, of course, states that the central energy density of
the maximum mass star is inversely proportional to the square of the
maximum mass.  Since the maximally compact EOS very closely predicts
the largest value for the maximum mass, it seems to follow that this
equation should also describe the maximum possible energy density as
well.  In other words, substituting the mass of the largest observed
neutron star mass for $M_{max}$ in the above expression not only
determines the largest energy density in that star, but it might yield
an approximate upper limit to the energy density to be found in {\it
any} neutron star.  Should a more massive star ever be measured, the
upper energy density limit would have to decrease.  

It is interesting that the analytic solution to the TOV equations for the
energy density distribution
\be
\varepsilon=\varepsilon_c[1-(r/R)^2],
\label{tolmanvii}
\ee 
{\it i.e.}, the Tolman VII solution~\cite{tolman39}, coupled with the
causality condition Eq. (\ref{maxc}), predicts a value for
$\varepsilon_{max}M_{max}^2$ to within a few percent of
Eq. (\ref{dm2}), even though the EOS resulting from
Eq. (\ref{tolmanvii}) becomes acausal at sufficiently large densities.
Ref.~\refcite{LP05} showed that the limit established from the Tolman
VII solution is satisfied phenomenologically by all realistic EOSs, as
illustrated in Fig. \ref{compfig1}.  By comparison, the limit defined
by Eq. (\ref{dm2}) is slightly violated by two of the EOS models but
nevertheless represents a realistic limit.  As was the case for the
estimate of the maximum neutron star mass, the slight violations
result from the assumption of zero pressure at low densities.
\begin{figure}[bt]
\centerline{\psfig{file=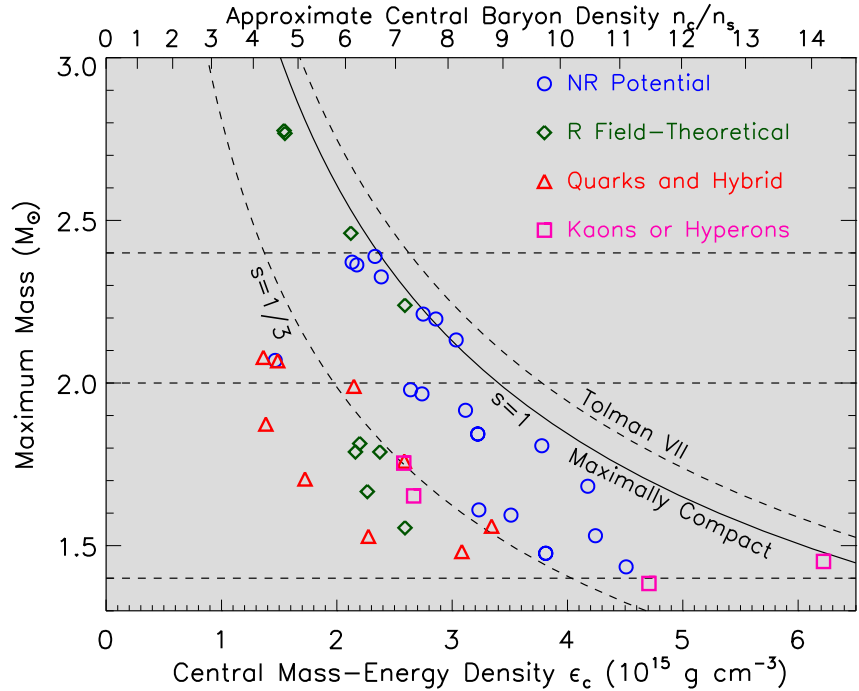,angle=90,width=5in}}
\caption{The central density versus mass relations for the maximally
compact EOS ($s=1$, Eq.~(\ref{dm2})) in comparison to Tolman VII + causality and the
$s=1/3$ (Eq.~(\ref{dm2q})) self-bound cases.  Various points are results from different
kinds of EOSs, as labelled.  Horizontal dashed lines indicate masses of
1.4, 2.0 and 2.4 M$_\odot$, respectively.}
\label{compfig1}
\end{figure}

Ref.~\refcite{KSF97} argues that the maximally compact EOS generally
predicts the smallest radius for a configuration with arbitrary mass.
In this event, the minimum compactness $R/M$ for any neutron star is
determined by
\be
R_{max}/M_{max}=2.824G/c^2\,
\label{maxc} 
\ee
a result consistent with discussions by Refs.~\refcite{Lindblom84,haensel89,LPMY90} and \refcite{Glendenning92}.  Modifying the EOS at low densities, i.e., for
$\varepsilon<(2-3)~\varepsilon_s$, does not significantly alter the
values for $M_{max}$, but the values for $R_{max}$ are moderately
increased~\cite{LPMY90}.

Ref. ~\refcite{KSF97} also demonstrated that the
axially symmetric structure equations in general relativity also scale
exactly for the maximally compact EOS in terms of $\epsilon_o$.  The
maximum rotation rate, limited by mass-shedding at the equator, scales
as $\sqrt{M_{max}/R_{max}^3}\propto\sqrt{\varepsilon_o}$ where the
mass and radius refer to the values for maximum mass of the
non-rotating configuration.  As a result, they calculated the minimum
spin period for a causally-limited EOS:
\be 
P_{min}=0.74\left({{\rm M}_\odot\over
M_{max}}\right)^{1/2}\left({R_{max}\over10{\rm~km}}\right)^{3/2}{\rm~ms}
=0.20{M_{max}\over{\rm M}_\odot}{\rm~ms}
\ee 
Lattimer and Prakash~\cite{LP04} showed that for realistic equations
of state the coefficient 0.74 ms is effectively replaced by 1.00 ms, a
value almost exactly equal to the general relativistic result in the
case that the gravitational potential of the rotating star is due to a
point mass at the origin~\cite{shapiro83}.  Furthermore, the revised
formula accurately predicts the maximum rotation rate of any star, not
just the maximum mass configuration, if $M_{max}$ and $R_{max}$ are
replaced by their non-rotating values.  

The maximally compact configuration is also expected to maximize the
binding energy,
\be {\rm
BE}_{max}=\left({m_bc^2\over\mu_o}z_{max}-y_{max}\right)
{c^6\over\varepsilon_o^{1/2}G^{3/2}}=0.337~M_{max}c^2.
\ee
The maximum binding energy is thus 25.2\% of the rest mass, a
fraction that is independent of $\varepsilon_o$ and $M_{max}$.

It is intersting that each of the cases shown in Table \ref{table1}
yields limiting values for $h_{max}$ and therefore $\mu_{max}$ that
are also independent of $\varepsilon_o$ and therefore also $M_{max}$.  In the
case of the maximally compact EOS, one finds the
upper limit to the baryon chemical potential
\be
\mu_{max}=2.251\times930~{\rm MeV} = 2093~{\rm  MeV}\,.  
\ee

These limits to the chemical potential, baryon number density, and
binding energy have not been previously calculated.  We next explore
the implications of the large neutron star masses recently measured, 
and whether the limits obtained from the ideal $s=1$ and $s=1/3$ cases
pose realistic upper bounds for thermodynamic quantities within neutron stars.

\section*{Implications of the newly-measured mass}

The measured mass of 1.97 M$_\odot$ can now be used to set 
specific limits because the value of $\epsilon_o$ is limited according to the first of Eqs. (\ref{mr}): $\varepsilon_o < 4.33~\varepsilon_s$.  This implies that
\be
\varepsilon_{max} &<& 3.034\times4.33~\varepsilon_s=13.1~\varepsilon_s=1.97{\rm~GeV~fm}^{-3}\,,\nonumber \\
p_{max} &<& 2.034\times4.33~\varepsilon_s=8.81~\varepsilon_s=1.32{\rm~GeV~fm}^{-3}\,,\nonumber \\ 
n_{max} &<& 2.251\times4.33~n_s=9.75~n_s=1.56{\rm~fm}^{-3}\,.
\label{emax}
\ee
Recall that the upper limit to the chemical potential, $\mu_{max}\simeq2.23\varepsilon_s/n_s= 2.1$ GeV, is
independent of the measured mass.

Finally, let us note the implications of a hypothetical measurement of
 2.4 M$_\odot$, for example, were the current estimate of the black
 widow pulsar to be confirmed.  In this case, by virtue of the first
 of Eqs. (\ref{mr}), we find $\varepsilon_o <2.92~\varepsilon_s$.  This in
 turn implies that
\be
\varepsilon_{max} &<& 3.034\times2.92~\varepsilon_s=8.86~\varepsilon_s=1.33{\rm~GeV~fm}^{-3}\,,\nonumber \\
p_{max} &<& 2.034\times2.92~\varepsilon_s=5.94~\varepsilon_s=0.89{\rm~GeV~fm}^{-3}\,,\nonumber \\ 
n_{max} &<& 2.251\times2.92~n_s=6.57~n_s=1.05{\rm~fm}^{-3}\,.
\label{emax1}
\ee

\section{Quark Matter EOSs}

As at sufficiently high density a quark-hadron transition is likely to
occur, it is reasonable to consider a quark matter EOS for the highest
densities that might exist in neutron stars.  In the case that quark
matter has a lower energy per baryon at zero pressure than iron
nuclei, a pure quark star is possible in which case the pressure
vanishes at non-zero baryon density~\cite{Witten84}.  This is an
example of a self-bound star.  The radii of these stars decrease with
decreasing mass in contrast to the case of normal neutron stars.  The
existence of a thin hadronic crust on such a star is possible, but
this would not affect considerations of the maximum mass or central
density, pressure or chemical potential of the star, although the
mass-radius relation for small masses is dramatically altered.

The term hybrid star refers to the case in which the star's interior
contains exotica such as hyperons, Bose condensates and /or quark
matter whereas the star's exterior is made up of hadronic matter as in
a normal neutron star.  The presence of matter extending to vanishing
density causes the radii of hybrid stars to increase with decreasing
mass as for normal neutron stars.  A quark matter EOS is obviously
softer than the maximally compact EOS, and since the maximum mass
depends primarily on the EOS at the highest densities, the maximal
constraints for pure quark stars or for neutron stars with quark cores
will be more severe than for the maximally compact case.  A relevant
question is whether or not the transition to quark matter is at low
enough density to occur in a neutron star.

\section*{MIT Bag Model}
Quark matter is likely to be highly interacting for densities near the
transition density.  Nevertheless, the leading order interaction terms
are generally attractive so that a reasonable upper limit to the
pressure of quark matter is that for non-interacting massless
quarks,
\be 
p = \frac 13 (\epsilon-4B) \,,
\label{q}
\ee
where $B$ is the MIT bag constant~(see, e.g., Ref. \refcite{FJ84}).
This is just the EOS of Eq.~(\ref{spec}) with $s=1/3$ and
$\epsilon_o=4B$. Even if QCD corrections are incorporated, the generic
form in Eq.~(\ref{spec}) is not greatly altered~\cite{Alford05}.
Additional corrections for a non-zero strange quark mass and for
pairing introduces attractive contributions~\cite{Alford05} and must
lower the maximum mass so we ignore these for now.  If the EOS of
quark matter approaches the ideal of Eq. (\ref{q}) at high density,
and there is significant quark matter in a neutron star interior, we
conjecture that limits observed in the case $s=1/3$ could more
realistically be expected to approximately constrain the maximum mass
and maximum possible energy density, pressure and chemical potential
than what is determined from the maximally compact EOS.
Irrespective of the value
for $s$, the relation $h=\ln(\mu/\mu_o)$ is still valid.  The limits
to mass and compactness were investigated for the case $s=1/3$ by
LPMY~\cite{LPMY90}.  The results for the dimensionless integrations
are contained in Table \ref{table1}.  All results for structural
variables scale with $\varepsilon_o$ just as in the maximally compact
case.

>From Table \ref{table1} it is clear for the case $s=1/3$ that the maximum
value of $h$, $h_{max}=0.4520$, is considerably smaller than its value for the
maximally compact EOS.  The maximum mass stars are also not as compact, the
limiting $R/M$ being 3.696.  The various dimensionless limits for the
case $s=1/3$ include
\be
M_{max} &=& 2.48~(\varepsilon_s/\varepsilon_o)^{1/2}{\rm~M}_\odot \,,\nonumber \\
N_{max}m_b &=& 2.98~(m_bc^2/\mu_o)(\varepsilon_s/\varepsilon_o)^{1/2}{\rm~M}_\odot \,,\nonumber \\
R_{max} &=& 13.56~(\varepsilon_s/\varepsilon_o)^{1/2}{\rm~ km} \,, \nonumber \\
\mu_{max} &=& \mu_oe^{h_{max}}=1.571~\mu_o \,,\nonumber \\ 
\varepsilon_{max} &=& (e^{2h_{max}}+1)\varepsilon_o/2=4.826~\varepsilon_o\,,\nonumber \\
p_{max} &=& (e^{2h_{max}}-1)\varepsilon_o=1.275~\varepsilon_o\,,\nonumber \\
n_{max} &=& (\varepsilon_{max}+p_{max})/\mu_{max}=
3.884~(\varepsilon_o/\mu_o) \,.
\label{mrq}
\ee
The results for $M_{max}, R_{max}$ and $\varepsilon_{max}$ were
previously obtained by Witten~\cite{Witten84}.  A result of great
significance is that, under the reasonable assumption that
$\varepsilon_o>\varepsilon_s$ (otherwise deconfined quark matter would be
visible in atomic nuclei), the maximum mass of a pure quark star
or a hybrid star with a substantial quark core can be no greater than
2.48 M$_\odot$.  We will see that, in the case of pure quark stars,
more realistic quark models generally have smaller pressures for a
given energy density than the MIT bag model and therefore smaller
maximum masses.  In the case of hybrid stars, if quark matter exists
in the cores, the limiting EOS is essentially that of pure quark
matter.
The magnitude of the transition density $n_t$ depends sensitively on
whether the hadronic EOS is soft or stiff at the relevant
densities. For stiff hadronic EOSs $n_t$ is often larger than the
central density of the maximum mass star and quarks will not appear.
In the case that $n_t$ is less than several times $n_s$, we will show
that the hadronic EOS at lower densities does not substantially alter
the maximum mass.  Thus, the observation of a 2.4 M$_\odot$ neutron
star, with the certainty that the true maximum mass is larger, will
greatly challenge the notion of the existence of quark matter in
neutron star interiors.

It is important to note that the chemical potential $\mu=\mu_o\ln h$
refers to the baryon chemical potential and $n$ refers to the number
density of baryons.  For neutral quark matter with three flavors, the
quark chemical potential is 1/3 the baryon (neutron) chemical
potential and the quark density is 3 times the baryon density.
Nevertheless, the value of $n\mu$ is the same for both types of
matter.  The value of $\mu_o$ will depend on assumptions regarding the
surface of quark matter stars.  It is widely believed that a thin
baryonic crust will exist, and in this case, the value of $\mu_o$ will
be the same as previously considered: Iron at zero pressure with
$\mu_o\simeq930$ MeV. If $\mu_o=930$ MeV, then we immediately find for
the case $s=1/3$ that $\mu_{max}=1.46$ GeV.  As we show below, this
limit for a self-bound MIT bag quark star remains approximately valid
for self-bound interacting quark stars and for hybrid quark stars with
a hadronic exterior.

Importantly, the TOV equations do
not depend on the value for $\mu_o$ as can be seen from
Eq.~(\ref{TOV2}), but absolute values for the number density and
chemical potential do depend on this quantity.
In order to discuss separate limits for the chemical potential or number
density, one has to have a specific model for the dependence of the
energy density on the chemical potential.  For 
massless quarks in the MIT bag model with bag constant $B$, the relation is
\be
\varepsilon={9\over4\pi^2(\hbar c)^3}\mu_q^4+B \,.
\label{MIT}
\ee
where $\mu_q=\mu/3$ is the quark chemical potential.  Each of the
three flavors of quarks has the same chemical potential and density in
the massless, neutral, case.  We can therefore directly
use the results from Table \ref{table1} using $\varepsilon_o=4B$ to obtain
\be
\mu_{qo} &=& \left({\epsilon_o\pi^2\over3}\right)^{1/4}(\hbar c)^{3/4}
= 268.6 \left({B^{1/4}\over141{\rm~MeV}}\right){\rm~MeV}\,,
\nonumber \\
n_{qo} &=& \left({\varepsilon+p\over\mu}\right)_o
= \left({3\over\pi^2}\right)^{1/4} \left({\epsilon_o\over\hbar c}\right)^{3/4}
=0.766 \left({B^{1/4}\over141{\rm~MeV}}\right)^3{\rm~fm}^{-3} \,. 
\label{muo}
\ee 
Note that the limiting value for $\mu_{qo}<310$ MeV fom the condition of
stability of quark matter at zero pressure implies $B^{1/4}<163$ MeV
for the MIT bag model~\cite{Witten84}.

\section*{Constraints from the 2-solar-mass star}
We now see that the observed mass of 1.97 M$_\odot$ suggests that
\be
\varepsilon_o&<&(2.48/1.97)^2\varepsilon_s=1.59~\varepsilon_s,\qquad
B^{1/4}<146.3{\rm~MeV} \,\nonumber \\ 
\mu_{qo}&<&\left({1.59\varepsilon_s\pi^2\over3}\right)^{1/4}(\hbar c)^{3/4}=278.7~{\rm~MeV}\,,
\nonumber \\ 
n_{qo}&<& \left({3\over\pi^2}\right)^{1/4} \left({1.59\varepsilon_s\over\hbar c}\right)^{3/4}=0.856~{\rm~fm}^{-3}\,.
\label{epc}
\ee
Therefore, the MIT bag model (without strong interaction corrections) cannot explain the 1.97 M$_\odot$ star
unless strange quark matter is stable.  The condition that
$\mu_{qo}>310$ MeV, necessary for iron to be the ground state of
matter at zero pressure, implies that the maximum mass is 1.58
M$_\odot$ if the MIT bag model is correct.

One also finds
\be
\varepsilon_{max} &=& 1.59\times4.826~\varepsilon_s=7.669~\varepsilon_s \,, \nonumber \\
p_{max} &=& 1.59\times1.275~\varepsilon_s=2.026~\varepsilon_s\,, \nonumber \\
\mu_{q,max} &=& 1.571\mu_{qo}=437.8{\rm~MeV}\,, \nonumber \\
n_{b,max} &=& {p_{max}+\varepsilon_{max}\over3\mu_{q,max}} =6.926~n_s\,,
\label{epc1}
\ee
and
\be
\varepsilon_{max}M_{max}^2 = 7.91\times10^{15}{\rm~g~cm}^{-3}{\rm~M}_\odot^2 \,.
\label{dm2q}
\ee
This relation is shown in Fig.~\ref{compfig1}.
The 1.97 M$_\odot$ star thus limits the energy density to
$2.0\times10^{15}$ g cm$^{-3}=1.13$ GeV fm$^{-3}$, just about half the
value for the maximally compact EOS. 
Likewise, the maximum
neutron chemical potential is $\mu_{max}=3\mu_{q,max}$ or $1.29$ GeV
which is considerably less than the absolute upper limit of 2.09 GeV
from causality alone.

Note that the reasonable constraint $\varepsilon_o>\varepsilon_s$
implies $B^{1/4}>130.3$ MeV in this model.  The relatively
narrow restrictions on the value of $B$ imply that
$248.2~{\rm MeV} < \mu_{qo} <278.7$ MeV.  If one further assumes that the
baryon density at the surface of the star is at $q$ times the nuclear
saturation density, or $n_{qo}=3qn_s$, one has
$1.26<q<1.78$.  If a neutron star of 2.4 M$_\odot$ were
to be found, this range would shrink to an extreme degree.

\section*{Modifications to the Bag Model}

In the model developed by Alford et al.~\cite{Alford05}, QCD
corrections are included with a parameter $a_4$, and the effects of a
finite strange quark mass and pairing are included with a term
proportional to the parameter $a_2$:
\be
\varepsilon={9a_4\over4\pi^2(\hbar c)^3}\mu_q^4-{3a_2\over4\pi^2(\hbar
c)^3}\mu_q^2+B_{eff}\,,
\label{MIT1}
\ee
where $B_{eff}$ is an effective bag
constant.
 
We first consider the case when $a_2$=0. Obviously, the introduction
of the parameter $a_4$ and the replacement of $B$ with $B_{eff}$ will
not affect the structure of the star and the dimensionless results
from the case $s=1/3$ still apply as for the MIT bag model.  The
introduction of the parameter $a_4$ affects the chemical potential and
number densities, however. The quark chemical potential and quark
number density at the surface of the star where $p=0$ can now be
determined for $a_2=0$ (see also Ref. \refcite{PBP90}):

\be
\mu_{qo} &=& \left({\epsilon_o\pi^2\over3a_4}\right)^{1/4}(\hbar c)^{3/4}
= 268.6 \left({B_{eff}^{1/4}\over141{\rm~MeV}}\right)a_4^{-1/4}~{\rm~MeV}\,,
\nonumber \\
n_{qo} &=& \left({\varepsilon+p\over\mu}\right)_o
= \left({3a_4\over\pi^2}\right)^{1/4} \left({\epsilon_o\over\hbar c}\right)^{3/4}
=0.766 \left({B_{eff}^{1/4}\over141{\rm~MeV}}\right)^3a_4^{1/4}~{\rm~fm}^{-3} \,. 
\label{muo2}
\ee

In this case the observed mass of 1.97 M$_\odot$ suggests that
\be
\mu_{qo}&<&\left({1.59\varepsilon_s\pi^2\over3a_4}\right)^{1/4}(\hbar c)^{3/4}=278.7~a_4^{-1/4}{\rm~MeV}\,,
\nonumber \\ 
n_{qo}&<& \left({3a_4\over\pi^2}\right)^{1/4} \left({1.59\varepsilon_s\over\hbar c}\right)^{3/4}=0.856~a_4^{1/4}{\rm~fm}^{-3}\,.
\label{epc2}
\ee
It is now evident that $a_4<0.656$ is required if both $\mu_{qo}=310$ MeV
and $M_{max}>1.97$ M$_\odot$ in this model.  Further, one has
\be
\mu_{q,max} &=& 1.571\mu_{qo}=437.8~a_4^{-1/4}{\rm~MeV}\,, \nonumber \\
n_{b,max} &=& {p_{max}+\varepsilon_{max}\over3\mu_{q,max}} =6.926~a_4^{1/4}n_s\,.
\label{epc3}
\ee
The introduction of the $a_4$ term does not alter the limiting energy
density inferred from the observed 1.97 M$_\odot$ star.  However, the
maximum baryon chemical potential $\mu_{max}=1.29/a_4^{1/4}$ GeV
increases moderately from the noninteracting quark case.
Nevertheless, the condition that $\mu_{qo}=310$ MeV does not change
the limiting value of $\mu_{max}=1.46$ GeV from that case.  We note that one
expects from theories of interacting quark matter that
$a_4\sim0.7$~\cite{Alford05}.
 
The constraint $\varepsilon_o>\varepsilon_s$ implies
$B_{eff}^{1/4}>130.3$ MeV in this model, unchanged by the $a_4$
parameter.  The restriction on the value of $B_{eff}$ from the
observed 1.97 M$_\odot$ star now implies that
$248.2<\mu_{qo}a^{1/4}<278.7$ MeV.  Finally, assuming the baryon
density at the surface of the star to be $qn_s$, one finds
$1.26<qa_4^{-1/4}<1.78$.

We next have to determine how modifying the quark EOS by introducing
the $a_2$ term changes these results.
The $a_2$ term is generally a small correction to the quark EOS.
In solving the TOV equation, we may once again scale energy and
pressure with $\varepsilon_o=4B_{eff}$, but there is an additional
parameter $a_2$.  Furthermore, $\varepsilon_o$
is no longer the value of $\varepsilon$ when $p=0$.  So the
introduction of $a_2$ is a moderate complication.  

Although there are now 3 parameters affecting the quark EOS, there is a 
degeneracy in the parameters as far as structural considerations 
are concerned. We write Eq.~(\ref{MIT1}) in the dimensionless form
\be
{\epsilon\over B_{eff}}&=&{9a_4\over4\pi^2}\left({\mu_q\over
B_{eff}^{1/4}}\right)^4 -{3\over4\pi^2} \left({a_2^{1/2}\over
B_{eff}^{1/4}}\right)^2\left({\mu_q\over B_{eff}^{1/4}}\right)^2 +1 \,,\nonumber \\
{p\over B_{eff}}&=&{3a_4\over4\pi^2}\left({\mu_q\over
B_{eff}^{1/4}}\right)^4 -{3\over4\pi^2}\left({a_2^{1/2}\over
B_{eff}^{1/4}}\right)^2\left({\mu_q\over B_{eff}^{1/4}}\right)^2 -1\,.
\label{MIT2}
\ee
Therefore it is only
necessary to consider as parameters the quantities $a_4$ and
$a_2^{1/2}/B_{eff}^{1/4}$ in order to establish scaled results for the
maximum mass and central pressure and energy density.  The value
of $\mu_{qo}/B_{eff}^{1/4}$ is a quadratic solution of the equation for $p=0$
from Eq.~(\ref{MIT2}).
To compute numerical values for the mass, energy density, pressure and
chemical potential, it is then additionally necessary to fix $B_{eff}$.  
However, this is not necessary to establish the quantity $\varepsilon_{max}y_{max}^2/\varepsilon_o$ which is independent of $\varepsilon_o=4B_{eff}$.

We find that varying $a_4$ for fixed nonzero $a_2^{1/2}/B_{eff}^{1/4}$
has small effects on eigenvalues $h_{max}, x_{max}$, $y_{max}$ and
$\varepsilon_{max}y_{max}^2/\varepsilon_o$.  Likewise, there are
similar small variations of these eigenvalues with varying
$a_2^{1/2}/B_{eff}^{1/4}$ for fixed values of $a_4$.  In any event,
the values of $h_{max}, x_{max}$ and
$\varepsilon_{max}y_{max}^2/\varepsilon_o$ are largest for the case
$a_4=1$ and $a_2=0$, as seen in Table \ref{table1}.  Furthermore, the
increase in the upper limit to the chemical potential resulting from
$a_4<1$ is approximately canceled by the $a_2$ contribution.
Therefore, the upper limits to the central density and pressure
established by the masssless quark bag model hold without regard to
QCD corrections, pairing, or finite strange quark mass.  This result
is confirmed phenomenologically, as illustrated in Fig. \ref{compfig1}
which indicates that the limiting density curve for $s=1/3$, to within
a few percent, bounds pure quark stars and hybrid stars containing
both quark and hadronic matter.  As we proceed to show, the addition
of a hadronic mantle and/or crust may permit densities slightly larger
than that predicted by Eq. (\ref{dm2q}), just as they increased the
densities for some hadronic neutron stars in some cases beyond the $s=1$
curve in Fig. \ref{compfig1}.

\section*{Simple Hybrid Stars}

The thermodynamic limits we have so far considered for quark matter
apply to self-bound configurations with non-zero baryon density at
their surface or those with a very thin hadronic crust.  Hybrid stars
consist of a quark core or a quark-hadron mixed phase together with a
massive hadronic exterior.  The presence of a mixed phase is possible
only if the surface tension of quark matter is sufficiently large.  We
consider here the case where there is no mixed phase and the
transition between quark and hadronic matter is a first-order
transition at the point where $\mu$ and $p$ in the two phases are
equal, with values $\mu_t$ and $p_t$, respectively.

We will employ a straightforward example.  A reasonable 
approximation for hadronic pressures in the vicinity of nuclear matter 
densities for neutron star matter is
$p=K\varepsilon^2$,
where $K\simeq 1.1\cdot10^{-4}$ fm$^{3}$ MeV$^{-1}$.  For $h>h_t$ we
will use the general form for self-bound matter, Eq. (\ref{spec}).
Requiring $\mu$ and $p$ to match at $h_t$, we find
\be
\varepsilon&=&(e^{h/2}-1)/K\hspace*{6cm}h<h_t\\
\varepsilon&=&\varepsilon_0/t+e^{(h-h_t)t}[\varepsilon_o/t+(e^{h_t/2}-1)^2/K]/s\hspace*{2cm} h>h_t.
\ee
In general, $\varepsilon$ and $n$ will be discontinuous at $h_t$, but 
we can require them to be continuous if we set
\be
\varepsilon_o={e^{h_t/2}-1\over K}\left[1-{e^{h_t/2}-1\over s}\right].
\ee
This assumption does not have a significant effect on our results, and
has the advantage of reducing the number of free parameters in this
model to a single one: $h_t$, or, equivalently, $p_t$.  In the case
that $h_t$ or $p_t$ tends to zero, we recover the self-bound cases.
$h_{max}$ increases slowly with $p_t$, approximately as $\ln(1+p_t/p_s)$.  For
the case $p_t=10p_s$, integration of the TOV equation with this EOS
gives values for the quantities in Table \ref{table1} in the rows
denoted $^{**}$.

It is observed in both $s=1$ and $s=1/3$ cases with added shells and
$p_t=10p_s$ that $h_{max}$ modestly increases by about 0.04, {\it
i.e.}, $\mu_{max}$ is increased by about 4\%.  Further increasing
$p_t$ results in marginally larger values of $h_{max}$.  Because the
self-bound cores in both the $s=1$ and $s=1/3$ cases are stiffer than
is realistic for hadronic and quark matter, respectively, which tends
to unrealistically enhance the values of $h_{max}$, we conclude that
the effects of adding a hadronic shell does not result in a
significant increase in the central chemical potential, and the values
established for the self-bound cases still hold approximately.
However, values of $\varepsilon_{max}y_{max}^2/\varepsilon_o$ decrease
with the addition of a hadronic shell, so our model is not
sufficiently general to explain the small violations of the $s=1$ and
$s=1/3$ curves seen in Fig. \ref{compfig1}.  We now compare these
results, in the case for quark stars, with those from published
models.

\section*{Published Models of Hybrid Stars}

>From many calculations in published literature it can be
concluded that the density at which exotica appear is 
uncertain and depends sensitively on the nature of nucleonic
interactions up to $(2-3)~n_s$.  In most cases, however, the EOS with
exotica is softer than that without.  Consequently, the maximum mass
that a hybrid star can support is often smaller (with few exceptions)
than that of a pure nucleonic star. In view of the observation of a
$1.97~{\rm M}_\odot$ star, the natural question to ask is whether or
not exotica are even permitted within such stars.

At the current stage, it is difficult to rule out exotica as many
model calculations indicate that a 2 ${\rm M}_\odot$ can indeed be
supported by hybrid stars of different varieties.  Indeed,
Fig. \ref{compfig1} contains examples of hybrid quark stars exceeding
this threshhold.  As another example, Fig.~5 of Ref.~\refcite{thorsson94}
shows kaon-condensed stars of 2 ${\rm M}_\odot$ that would be
compatible with the current observation. Ref.~\refcite{SPL00}, which
contrasts the MIT bag model with the Nambu Jona-Lasinio model for the
treatment of quarks, presents results of hybrid stars containing
nucleons and quarks, and nucleons, hyperons and quarks (see Figs.~4 of
Ref. \refcite{SPL00}) with maximum masses of $\sim 2 {\rm M}_\odot$. The
central densities of maximum mass configurations obtained there were
in the range $(0.7-0.75)~{\rm fm}^{-3}$ (see Fig.~5 of
Ref.~\refcite{SPL00}), which is compatible with the limiting curve
$s=1/3$ in Fig. \ref{compfig1}.  It would be interesting to see whether or not Dyson-Schwinger -equation approaches~\cite{Klahn10} to the quark matter EOS yield similar results.   

Recently, perturbative calculations of quark matter (without effects
of pairing) including strong interaction effects to order $\alpha_s^2$
and the effects of the strange quark mass have been performed by
Kurkela et al. in Ref.~\refcite{KRV10}.  As these authors point out, their
results become reliable at the $10\%$ level only for quark chemical
potentials exceeding 1 GeV, a value far larger than can exist in pure
quark or quark hybrid stars as we have shown (the maximum is about
0.5 GeV). Their calculations nevertheless result in quark hybrid stars
with masses approaching $2~{\rm M}_\odot$.  

These examples show that it is at present not possible to rule out any
form of exotica with the 1.97 M$_\odot$ observation. It must be
stressed, however, that many of these approaches suffer from a great
deal of model dependence (even in cases in which a substantially lower
than a $2~{\rm M}_\odot$ maximum mass star is predicted).  However,
the situation would change dramatically in the event of a confirmed
observation of a 2.4 M$_\odot$ star.  Such an observation would seem to be
incompatible with quark or quark hybrid stars including interactions,
and would severely restrict models containing other
forms of exotica.

\section*{Gerry's Views on Hybrid Stars}

Prior to his fascination with kaon condensates, Gerry, together with
Hans Bethe and Jerry Coopertsein~\cite{BBC87}, had maintained that
perturbative QCD cannot be trusted at densities near $n_s$ where
matter must be confined, and that any strangeness would appear in the
form of Lambdas.  To infer whether strangeness-rich quark matter
can occur at high densities, they calculated the energy density of
massless up, down and strange quarks to order $\alpha_s$ taking its
density dependence as $\alpha_s=2.2(n/n_s)^{1/3}$.  However, as shown
in Ref. \refcite{PBP90}, their choice turned out to be independent of
$\alpha_s$ once thermodynamic consistency was imposed. The result of
their calculations was that the transition to quark matter occurred at
densities larger than those found in neutron stars.

A lesson lurks in Gerry's point of view. Most models of hyperon-filled
stars have used hyperon-nucleon couplings based on simplistic quark
counting in the non-perturbative regime.  Analyses of laboratory
experiments, however, indicate that at nuclear densities the
$\Lambda$-nucleon potential is attractive, but the $\Sigma^-$-nucleon
potential is repulsive~\cite{Friedman94,Mares95}.  Knorren, Prakash
and Ellis~\cite{KPE95} investigated the consequences of varying
hyperon-nucleon couplings on the structural properties of neutron
stars and concluded that, if only Lambdas were to exist in neutron
stars, their effects on the maximum mass would be minimal. If such is
indeed the case, then two birds can be killed with one stone: first, a
star with a 2-solar mass could be easily attained, and second, an
avenue for a fast cooling could arise even if the direct Urca process
with nucleons could not take place as even miniscule amounts of
Lambdas through their beta decays can cool the stars
rapidly~\cite{PPLP92}.  In a final example, based on a field-theoretical
model for the hadrons and an NJL model for the quarks,
Ref.~\refcite{SGMT07} fits observed $\Lambda$ and $\Xi^-$ hyper-nuclei
and is able to construct hybrid stars of 2 M$_\odot$.  Further
investigations to fortify these ideas appear promising.
           
\section{Neutron Matter and Astrophysical Constraints}

It is also possible to place constraints on the maximum mass by
combining the condition of causality with laboratory data concerning
the symmetry energy, neutron skins on neutron-rich nuclei, and the
properties of pure neutron matter as well as with astrophysical
measurements of neutron star radii and masses.  Recently, estimates of
neutron matter properties have been obtained by first principle
Green's Function Monte Carlo calculations \cite{carlson} and from
chiral effective interactions \cite{HS}.  These approaches yield
similar results.  Hebeler et al. \cite{HLPS} showed that such
calculations restrict neutron star radii to be in the range of (11-12)
km when the star's central density is at a density of $1.1n_s$,
roughly the uppermost density for which the neutron matter
calculations are reliable.  With this constraint, it was possible to
show that causality limits the maximum mass to about 2.9 M$_\odot$
(see Fig. 3 of Ref. \refcite{HLPS}).

Steiner, Lattimer and Brown \cite{SLB} demonstrated that estimates of
masses and radii of neutron stars in photospheric radius expansion
X-ray bursters and also quiescent, cooling neutron stars limit the
radii of 1.5 M$_\odot$ stars to lie in the radius range (10.9-12.5)
km.  This radius range resulted from a Bayesian analysis of the
observational data from three sources of both types.
Encouragingly, this radius range is also consistent with
the results of neutron matter simulations~\cite{carlson,HLPS}.  With
this mass-radius constraint, causality limits the maximum mass to
about 2.4 M$_\odot$, about 0.5 M$_\odot$ smaller than that inferred
from the neutron matter results themselves.  This upper limit too the
maximum mass is due to the
fact that the radius does not rapidly change with mass for a
significant part of the mass-radius relation, as can be observed in
Fig. \ref{tovfig}.  Thus, a restriction of the radius at a specific mass
translates into an upper mass limit when the causality condition ({\it i.e.},
$M<Rc^2/(2.9G)$) is imposed.

\section{Conclusions}

Based on the observed 
largest well-measured mass, we have determined model-independent upper
limits to the central energy density, central pressure and central
chemical potential of neutron stars whether or not they are composed
of hadrons, quarks, or both.  If matter undergoes a quark-hadron
transition within the density range permitted for neutron star
interiors, these limits on central density and chemical potential are
substantially lowered.  Indeed, if it can be shown that the maximum
mass of neutron stars exceeds 2.5 M$_\odot$, the possibility of a
quark-hadron transition in neutron stars is actually excluded, and the
transition in cold matter would have to occur at an energy density in
excess of $8\varepsilon_s$ or about 1.2 GeV.

We have also demonstrated the existence of an upper limit to the
baryon chemical potential, independent of mass observations, of 2.1
GeV.  If quark matter exists in the cores of neutron stars, the upper
limit is likely not greater than 1.5 GeV.  We also find the maximum
binding energy or any neutron star is about 25\% of the rest mass.

The astronomical evidence currently favors the neutron
star maximum mass to be between 2 M$_\odot$ and 2.5 M$_\odot$.
A Bayesian analysis of the observed neutron star mass distribution
suggests that this distribution shows no indication of being
influenced by an absolute upper limit imposed by general relativity
and causality \cite{kiziltan10}.  In other words, this distribution
shows no evidence of the existence of an abrupt upper mass cutoff;
rather, the largest observed mass may simply be limited by the amount
of mass it is reasonably possible for a neutron star to accrete in a
binary.  Supernova models generally imply that newly formed neutron
stars are not more massive than about 1.5 M$_\odot$.  
In this case it is likely that the true maximum mass
determined by the EOS is at least a few tenths of a solar mass larger than the
largest observed mass.  If so, the chemical potential, density and
pressure limits we derived are somewhat conservative.  In addition, our survey
of EOSs with exotic matter indicates that stars
containing {\it significant} proportions of this matter in the form of
hyperons, Bose condensates or quark matter would be almost impossible.

\section*{Acknowlegements}
This research was supported by the Department of Energy under the
grants DE-AC02-87ER40317 (for JML) and DE-FG02-93ER40756 (for MP).  We
also value our many discussions with Gerry on topics related to the neutron
star maximum mass.


\section*{References}

\begin{thebibliography}{0}
%
\bibitem{KN86} D. Kaplan and A. Nelson, {\it Phys. Lett. } {\bf B175}, 57 (1986); 
{\bf B179}, 409 (1986).   

\bibitem{brown94}
G. E. Brown and H. A. Bethe, {\it Astrophys. J.} {\bf423}, 659 (1994).

\bibitem{thorsson94} 
V. Thorsson, M. Prakash and J. M. Lattimer, {\it Nucl. Phys. A}  {\bf 572}, 693 (1994).
  
\bibitem{Burrows88} A. Burrows, {\it Astrophys. J.}  {\bf 334}, 891 (1988). 

\bibitem{manchester1977} R. N. Manchester and J. H. Taylor, Pulsars
(W. H. Freeman, San Francisco, 1977).

\bibitem{Shapiro64} I. I. Shapiro, {\it Phys. Rev. Lett.} {\bf 26}, 789 (1964).

\bibitem{damour86} T. Damour and N. Deruelle, {\it Ann. Inst. Henri
  Poincar\'e Phys. Theor.} {\bf44}, 263 (1986).

\bibitem{freirewex10} P. C. C. Freire and N. Wex, {\it MNRAS}
  {\bf409}, 199 (2010).

\bibitem{reynolds2007}
  M. T. Reynolds, P. J. Callanan, A. S. Fruchter, M. A. P. Torres,
  M. E. Beer and R. A. Gibbons, {\it MNRAS} {\bf379}, 1117 (2007).

\bibitem{clark02} 
J.S. Clark, S. P. Goodwin, P. A. Crowther, L.  Kaper, M. Fairbairn, 
N. Langer, and C. Brocksopp, 
{\it Astron. \& Astrophys.} {\bf392}, 909 (2002).

\bibitem{barziv01}
O. Barziv, L. Karper, M.H. van Kerkwijk, J.H. Telging, and J. van Paradijs,
{\it Astron. \& Astrophys.} {\bf377}, 925 (2001).

\bibitem{quaintrell03}
H. Quaintrell, A.J. Norton, T.D.C. Ash, P. Roche, B. Willems, 
T.R. Bedding, I.K. Baldry and R.P. Fender,
{\it Astron. \& Astrophys.} {\bf401}, 303 (2003).


\bibitem{casares10}
J. Casares, J.I. Gonz\'alez Hern\'andez,
  G. Israelian and R. Rebolo, {\it MNRAS} {\bf401}, 2517 (2010).

\bibitem{vankerkwijk95}
M.H. van Kerkwijk, J. van Paradijs, 
and E.J. Zuiderwijk, {\it Astron. \& Astrophys.} {\bf303}, 497 (1995).

\bibitem{vandermeer07}
A.van der Meer, L. Kaper, M. H. van Kerkwijk, 
M. H. M. Heermskerk and E. P. J. van den Huevel, 
{\it Astron. \& Astrophys.} {\bf473}, 523 (2007). 

\bibitem{gelino03}
D.M. Gelino, J.A. Tomsick and W.A. Heindl, 
{\it Bull. Am. Astron. Soc.} {\bf34}, 1199 (2003).

\bibitem{tomsick04}
J. Tomsick, private communication (2004).

\bibitem{steeghs07}
D. Steeghs and P. G. Jonker, {\it ApJLetters} {\bf669}, 85 (2007).

\bibitem{jonker03}
P. G. Jonker, M. van der Klis, P. J. Groot, 
{\it MNRAS} {\bf339}, 663 (2003).

\bibitem{champion05}
D.J. Champion, D. R. Lorimer, M.A. McLaughlin, K.M. Xilouris, 
Z. Arzoumanian, P.C.C. Freire, A. N. Lommen, J.M. Cordes and F. Camilo, 
{\it MNRAS} {\bf363}, 929 (2005).

\bibitem{corongiu04}
A. Corongiu , M. Kramer, A.G. Lyne, O. L\"ohmer, N. D'Amico, \& 
A. Possenti, {\it Mem. S. A. It Suppl.} {\bf5}, 188 (2004).

\bibitem{lorimer06}
D. R. Lorimer et al.,  {\it ApJ} {\bf 640}, 428 (2006).

\bibitem{janssen08}
G.H. Janssen, B.W. Stappers, M. Kramer, D.J. Nice, A. Jessner, 
I. Cognard and M.B. Purver, 
{\it Astron. \& Astrophys.} {\bf490}, 753 (2008).

\bibitem{stairs02}
I.H. Stairs, S.E. Thorsett, J.H. Taylor and A. Wolszczan, 
{\it ApJ} {\bf581} 501 (2002).

\bibitem{weisberg10}
J. M. Weisberg, D. J. Nice \& J. H. Taylor, {\it ApJ} {\bf722}, 1030 (2010).

\bibitem{jacoby06}
B.A. Jacoby, P.B. Cameron, F.A. Jenet, S.B. Anderson, R.N. Murty and 
S.R. Kulkarni, {\it ApJLetters} {\bf644}, L113 (2006).

\bibitem{kramer06}
M. Kramer, I.H. Stairs, R.N. Manchester, M.A. McLaughlin, A.G. Lyne, 
R.D. Ferdman, M. Burgay, D.R. Lorimer, A. Possenti, N. D'Amico, 
J.M. Sarkisian, G.B. Hobbs, J.E. Reynolds, P.C.C. Freire and F. Camilo, 
{\it Science} {\bf 314}, 97 (2006).

\bibitem{ferdman08}
R.D. Ferdman, Ph.D. thesis, Univ. of British Columbia (2008).

\bibitem{thorsett99}
S.E. Thorsett and D. Chakrabarty, {\it Astrophys. J.} {\bf512},
288 (1999). 

\bibitem{lange01}
Ch. Lange, F. Camilo, N. Wex, M. Kramer, D.C. Backer, A.G. Lyne and
O. Doroshenko, {\it MNRAS} {\bf 326}, 274 (2001).

\bibitem{splaver05}
E.M. Splaver, D.J. Nice, I.H. Stairs, A.N. Lommen and D.C. Backer,
{\it ApJ} {\bf620}, 405 (2005).

\bibitem{nice03b}
D.J.  Nice, E.M. Splaver and I.H. Stairs, 
in {\it Radio Pulsars},
ed. M. Bailes, D. J.  Nice, and S. E. Thorsett
(Ast. Soc. Pac. 302, San Francisco, 2003). 

\bibitem{nice08}
D.J. Nice, I.H. Stairs and L.E. Kasian, in 
{\it 40 YEARS OF PULSARS: Millisecond Pulsars, Magnetars and More}, 
AIP Conference Proceedings {\bf 983}, 453 (2008).

\bibitem{verbiest08}
J.P.W. Verbiest, M. Bailes, W. van Straten, G.B. Hobbs, R.T. Edwards, 
R.N. Manchester, N.D.R. Bhat, J.M. Sarkissian, B.A. Jacoby and S.R. Kulkarni, 
{\it ApJ} {\bf679}, 675 (2008).

\bibitem{bhat08}
N.D.R. Bhat, M. Bailes and J.P.W. Verbiest, {\it Phys. Rev. D} {\bf77}, 
124017 (2008).

\bibitem{ransom05}
S. M. Ransom, J.W.T. Hessels, I.H. Stairs, P.C. Freire, F. Camilo, 
V.M. Kaspi and K.L. Kaplan, {\it Science} {\bf307}, 892 (2005).

\bibitem{hotan06}
A.W. Hotan, M. Bailes and S.M. Ord, {\it MNRAS} {\bf369}, 1502 (2006).

\bibitem{freire03}
P. C. C. Freire, F. Camilo, M. Kramer, D.R. Lorimer, A.G. Lyne, 
R.N. Manchester \& N.D'Amico, {\it MNRAS} {\bf340}, 1359 (2003).

\bibitem{freire07}
P.C.C. Freire, S.M. Ransom and Y. Gupta, {\it ApJ} {\bf662}, 1177 (2007).

\bibitem{freire08a}
P.C.C. Freire, A. Wolszcan, M. van den Berg and J.W.T. Hessels, 
{\it ApJ} {\bf679}, 1433 (2008).

\bibitem{freire08b}
P.C.C. Freire, S.M. Ransom, S. B\'egin, I.H. Stairs, J.W.T. Hessels, L.H. Frey
and F. Camilo, {\it ApJ} {\bf675}, 670 (2008).

\bibitem{freire08c}
P.C.C. Freire, B.A. Jacoby and M. Bailes, in {\it 40 YEARS OF PULSARS: 
Millisecond Pulsars, Magnetars and More}, 
AIP Conference Proceedings {\bf 983}, 488; arXiv:0711.1880 (2008). 

\bibitem{bassa06}
C.G. Bassa, M. H. van Kerkwijk, D. Koester and F. Verbunt, {\it Astron. \& Astrophys.} {\bf456}, 295 (2006).

\bibitem{nice01}
D. J. Nice, E. M. Splaver and I. H. Stairs, {\it ApJ} {\bf 549}, 516 (2001).

\bibitem{nice03a}
D.J. Nice, {\it IAU Proceedings}, Sydney, Australia (2003).

\bibitem{freire10} P. C. C. Freire, C. G. Bassa, N. Wex, I. H. Stairs,
D. J. Champion, S. M. Ransom, P. Lazarus, V. M. Kaspi,
J. W. T. Hessels, M. Kramer, J. M. Cordes, J. P. W. Verbiest,
P. Podsiadlowski, D. J. Nice, J. S. Deneva, D. R. Lorimer,
B. W. Stappers, M. A. McLaughlin and F. Camilo, {\it MNRAS}, in press;
arXiv:1011.5809 (2010).

\bibitem{mason10}
A. B. Mason, A. J. Norton, J. S. Clark, I. Negueruela \& P. Roche, {\it Astron. \& Astrophys.}
{\bf509}, 79 (2010). 

\bibitem{demorest10} 
P. B. Demorest, T. Pennucci, S. M. Ransom, M. S. E. Roberts and
 J. W. T. Hessels, {\it Nature}, {\bf 467}, 1081 (2010).

\bibitem{vankerkwijk10} M. H. van Kerkwijk, R. Breton \&
  S. R. Kulkarni, submitted to {\it ApJ} (2010); arXiv:1009.5427.

\bibitem{ferdman10}
R.D. Ferdman, I.H. Stairs, M. Kramer,
M.A. McLaughlin, D.R. Lorimer, D.J. Nice, R.N. Manchester, G. Hobbs,
A.G. Lyne, F. Camilo, A. Possenti, P.B. Demorest, I. Cognard,
G. Desvignes, G. Theureau, A. Faulkner, D.C. Backer, {\it ApJ}
{\bf711}, 764 (2010).

\bibitem{tauris99}
T. M. Tauris and G.J. Savonije, {\it Astron. \& Astrophys.} {\bf350}, 928 (1999).

\bibitem{LP05} J. M. Lattimer and M. Prakash, {\it Phys. Rev. Lett.}
  {\bf 94}, 111101 (2005).

\bibitem{TOV} R. C. Tolman, {\it Proc. Nat. Acad. Sci. U. S. A.,} {\bf
  20}, 3 (1934); J. R. Oppenheimer and G. M. Volkoff, {\it Phys. Rev.}
  {\bf 55}, 374 (1939).

\bibitem{RR74} C. E. Rhoades Jr. and R. Ruffini, 
{\it Phys. Rev. Lett.} {\bf 32}, 324 (1974).

\bibitem{haensel89} P. Haensel and J. L. Zdunik, {\it Nature}
{\bf340}, 617 (1989).

\bibitem{KSF97} S. Koranda, N. Stergioulas and J. L. Friedman, 
{\it ApJ} {\bf 488}, 799 (1997).

\bibitem{Witten84} E Witten, {\it Phys. Rev. D} {\bf 30}, 272 (1984).

\bibitem{LPMY90} J. M. Lattimer, M. Prakash. D. Masak and A. Yahil, 
{\it ApJ} {\bf 355}, 241 (1990).  

\bibitem{Lindblom84} L. Lindblom, {\it ApJ} {\bf278}, 364 (1984).   

\bibitem{Glendenning92} N. K. Glendenning, 
{\it Phys. Rev. D}  {\bf 46}, 4161 (1992).

\bibitem{LP04} J. M. Lattimer and M. Prakash, {\it Science} {\bf304},
536 (2004).

\bibitem{shapiro83} S. L. Shapiro, S. A. Teukolsky and I. Wasserman,
{\it ApJ} {\bf272}, 702 (1983).

\bibitem{tolman39} R. C. Tolman, Phys. Rev. {\bf 55}, 364 (1939).

\bibitem{FJ84} E. Fahri and R. L. Jaffe, Phys. Rev. {\bf D30},  2379 (1984).

\bibitem{Alford05} M. G. Alford, M. Braby, M. Paris and S. Reddy, 
{\it ApJ} {\bf 629}, 969 (2005).  

\bibitem{PBP90} Manju Prakash, E. Baron and M. Prakash,
{\it Phys. Lett. B} {\bf 243}, 175 (1990).

\bibitem{SPL00} A. W. Steiner, M. Prakash and J. M. Lattimer,
{\it Phys. Lett.} {\bf 486}, 239 (2000).

\bibitem{Klahn10} T. Kl\"ahn, C. D. Roberts, L. Chang, H. Chen and Yu-Xin Liu,
{\it Phys. Rev. C} {\bf 82}, 035801 (2010).

\bibitem{KRV10} A. Kurkela, P. Romatschke and A. Vuorinen,
{\it Phys. Rev. D} {\bf 81}, 105021 (2010). 

\bibitem{BBC87} H. A. Bethe, G. E. Brown and G. E. Brown, 
{\it Nucl. Phys. A} {\bf 467}, 791 (1987).

\bibitem{Friedman94}  E. Friedman, A. Gal,  and C. J. Batty,
{\t Nucl. Phys. A} {\bf 579}, 578 (1994).

\bibitem{Mares95} J. Mares, W. Friedman, A. Gal and B. K. Jennings,
{\it Nucl. Phys. A} {\bf 594}, 311 (1995).

\bibitem{KPE95} R. Knorren, M. Prakash and P. J. Ellis,
{\it Phys. Rev. C} {\bf 52}, 3470 (1995).

\bibitem{PPLP92} M. Prakash, Manju Prakash, J. M. Lattimer and C. J. Pethick, 
{\t Astrophys. J. Lett.} {\bf 390}, L77 (1992).

\bibitem{SGMT07} J. R. Stone, P. A. M. Guichon, H. M. Matevosyan and A. W. Thomas,
{\it Nucl. Phys.}, A 792, 341 (2007).

\bibitem{carlson} A. Gezerlis and J. Carlson, {\it Phys. Rev. C}
{\bf81}, 025803 (2010).

\bibitem{HS} K. Hebeler and A. Schwenk, {\it Phys. Rev. C} {\bf82},
014304 (2010).

\bibitem{HLPS} K. Hebeler, J. M. Lattimer, C. J. Pethick and
A. Schwenk, {\it Phys. Rev. Lett.} {\bf 105}, 161102 (2010).

\bibitem{SLB} A. W. Steiner, J. M. Lattimer and E. F. Brown, {\it ApJ} {\bf722}, 33 (2010).

\bibitem{kiziltan10} B. Kiziltan, A. Kottas and S. E. Thorsett,
submitted to ApJ; arXiv:1011.4291 (2010).


\end{thebibliography}
\end{document}